\documentclass[letterpaper, 10 pt, conference]{ieeeconf}  % Comment this line out
\IEEEoverridecommandlockouts                              % This command is only
\usepackage{graphicx}
\usepackage{float}
\usepackage{booktabs}
\usepackage{array}
\usepackage{amsmath} % assumes amsmath package installed
\usepackage{url}
\usepackage{amsmath}
\makeatletter
\def\maketag@@@#1{\hbox{\m@th\normalfont\normalsize#1}}
\makeatother
\usepackage{algorithm}
\usepackage[noend]{algpseudocode}
\usepackage{setspace}
\usepackage[keeplastbox]{flushend}
\usepackage{tikz}
\usepackage{lipsum}
\usepackage{comment}
\usepackage{fancyhdr}
\fancypagestyle{firstpage}{%
  \lhead{\vspace{-0.75 cm} \small \textbf{57th IEEE Conference on Decision and Control (CDC)\\December 17-19, 2018, Miami Beach, FL, USA}}
}

\title{\Large \bf
Real-Time Model Predictive Control for Energy Management \\ in Autonomous Underwater Vehicle 
}

\author{Niankai Yang$^{1}$, Mohammad Reza Amini$^{1}$, Matthew Johnson-Roberson$^{1}$ and Jing Sun$^{1}$% <-this % stops a space
\thanks{$^{1}$Niankai Yang, Mohammad Reza Amini, Matthew Johnson-Roberson and Jing Sun are with Department of Naval Architecture and Marine Engineering, University of Michigan, Ann Arbor, MI 48109 USA. (e-mail: \{ynk,\;mamini,\;mattjr,\;jingsun\}@umich.edu)}%
}
\begin{document}

\maketitle
\thispagestyle{firstpage}
\pagestyle{empty}

\makeatletter
\AfterEndEnvironment{algorithm}{\let\@algcomment\relax}
\AtEndEnvironment{algorithm}{\kern2pt\hrule\relax\vskip3pt\@algcomment}
\let\@algcomment\relax
\newcommand\algcomment[1]{\def\@algcomment{\footnotesize#1}}
\renewcommand\fs@ruled{\def\@fs@cfont{\bfseries}\let\@fs@capt\floatc@ruled
  \def\@fs@pre{\hrule height.8pt depth0pt \kern2pt}%
  \def\@fs@post{}%
  \def\@fs@mid{\kern2pt\hrule\kern2pt}%
  \let\@fs@iftopcapt\iftrue}
\makeatother

%%%%%%%%%%%%%%%%%%%%%%%%%%%%%%%%%%%%%%%%%%%%%%%%%%%%%%%%%%%%%%%%%%%%%%%%%%%%%%%%%%%%
\begin{abstract}

Improving endurance is crucial for extending the spatial and temporal operation range of autonomous underwater vehicles (AUVs). Considering the hardware constraints and the performance requirements, an intelligent energy management system is required to extend the operation range of AUVs. This paper presents a novel model predictive control (MPC) framework for energy-optimal point-to-point motion control of an AUV. In this scheme, the energy management problem of an AUV is reformulated as a surge motion optimization problem in two stages. First, a system-level energy minimization problem is solved by managing the trade-off between the energies required for overcoming the positive buoyancy and surge drag force in static optimization. Next, an MPC with a special cost function formulation is proposed to deal with transients and system dynamics. A switching logic for handling the transition between the static and dynamic stages is incorporated to reduce the computational efforts. 
Simulation results show that the proposed method is able to achieve near-optimal energy consumption with considerable lower computational complexity.

\end{abstract}

%%%%%%%%%%%%%%%%%%%%%%%%%%%%%%%%%%%%%%%%%%%%%%%%%%%%%%%%%%%%%%%%%%%%%%%%%%%%%%%
%%%% Two paragraphs are removed in introduction
%%%%%%%%%%%%%%%%%%%%%%%%%%%%%%%%%%%%%%%%%%%%%%%%%%%%%%%%%%%%%%%%%%%%%%%%%%%%%%%%%%%%
\section{INTRODUCTION}

Autonomous underwater vehicles (AUVs) are advancing the state-of-the-art in numerous oceanography and aquatic environmental monitoring applications. With high level of autonomy, AUVs are well-suited for deep water or long-range explorations. Endurance is a key consideration for AUV design and operation. With constraints on the construction cost and other competing design requirements, the on-board energy storage of AUV is often limited, making AUV endurance a challenge and priority for design and operation.

%Several solutions have been proposed in the literature to improve AUV endurance. The first approach is mainly focused on the AUV design modification. It has been shown in~\cite{eriksen2001seaglider,webb2001slocum} that by taking advantage of the hydrodynamic lift generated by buoyancy control, underwater gliders can be designed and deployed. \cite{santhakumar2013power} showed that optimized design of the thrusters leads to reduction in energy consumption during station keeping. In another study~\cite{hasvold2004clipper}, a high energy density battery was utilized to extend the operation range of an AUV. However, these elaborate design modification techniques either increase the construction cost or restrict the application of the vehicle.

%The second approach focuses on 

Extensive efforts were made to improve the energy consciousness of AUV control in either the planning stage or the execution stage. At the planning stage, the ocean currents, which can be comparable in magnitude to the average vehicle operational speed, are considered to obtain the optimal trajectories associated with the minimum energy consumption~\cite{subramani2015stochastic}. 
%In \cite{garau2005path}, $A^*$ algorithm was used to compute a horizontal energy-minimum path in 2D ocean environment with eddies in different intensities and scales by assuming that the energy cost is equivalent to the traveling time. 
%Genetic algorithm was used in~\cite{alvarez2004evolutionary} to calculate the optimal path in a strong space-time varying current environment using an oversimplified vehicle kinematic model for predicting the energy consumption. More elaborated dynamic models have also been exploited to achieve high energy efficiency in AUV operation. For example, 
In \cite{subramani2017energy}, a stochastic optimization method integrated with data-driven ocean modeling is presented to generate an energy-optimal trajectory with reference speed and heading. The Pontryagin's maximum principle (PMP) was applied in \cite{chyba2009increasing} to obtain the global optimal energy trajectory. However, the intensive computation associated with the above planning algorithms makes their real-time implementation infeasible. Moreover, since the planning is performed offline, the results are sensitive to uncertainties in the AUV model and time-varying disturbances. 

Works have also been reported on the design of trajectory tracking controllers for energy consumption reduction at the execution level. In \cite{sarkar2016modelling}, a sliding mode controller was designed with Euler-Lagrange based classical optimal control for minimizing the control efforts and saving energy. State-dependent Riccati equation was applied in \cite{geranmehr2015nonlinear} for an energy-efficient AUV tracking control. These control strategies assume a given planned trajectory and achieve energy savings by compromising the tracking accuracy. 

In this paper, we propose an execution-level controller for a point-to-point motion control of an AUV, while optimizing the vehicle trajectory online by utilizing dynamic vehicle model to improve the energy efficiency of the system. To this end, an innovative model predictive control (MPC) %\textcolor{red}{\cite{rawlings2009model}} \textcolor{blue!50}{(you don't need to cite any ref here, it is given and clear)} 
%%%%%%%%% this is the response to the first comment of reviewer 3
framework with low computational complexity is developed to facilitate real-time implementation. The main contribution of this paper relies on reformulating the energy management problem of an AUV as a surge motion optimization problem, which incorporates the heave-related energy spending as a terminal cost in MPC to capture the ``cost-to-go''. In order to reduce the computational efforts, the surge motion optimization is performed in static and dynamic stages via a switching strategy in the proposed MPC.

%The remainder of this paper is organized as follows: in Section \ref{section.2}, the simulation model of an AUV is introduced. The formulation of surge motion optimization problem and the development of control-oriented model are described in Section \ref{section.3}. The static optimization of the surge motion is illustrated in Section \ref{section.4}. Section \ref{section.5} presents the dynamic surge motion optimization. The overall real-time energy-optimal MPC framework is proposed in Section \ref{section.6}. Simulation results and discussions are presented in Section \ref{section.7}. Finally, conclusion and future works are discussed in Section \ref{section.8}. 

%%%%%%%%%%%%%%%%%%%%%%%%%%%%%%%%%%%%%%%%%%%%%%%%%%%%%%%%%%%%%%%%%%%%%%%%%%%%%%%%%%%% 
%%%%%%%%%%%%%%%%%%%%%%%%%%%%%%%%%%%%%%%%%%%%%%%%%%%%%%%%%%%%%%%%%%%%%%%%%%%%%%%%%%%%
\section{AUV Simulation Model} \label{section.2} 

In this study, the DROP-Sphere \cite{Iscar:2018aa}, a low cost, 6000m rated AUV designed for optical benthic mapping of the deep sea  is used as the virtual testbed for algorithm development. \vspace{-0.55cm}
%The physical characteristics of $Sphere$ will be first described, followed by the mathematical model. \vspace{-0.1cm}
\subsection{DROP-Sphere Configuration}
\begin{figure}[!h]
\centering
\includegraphics[width=2.7in]{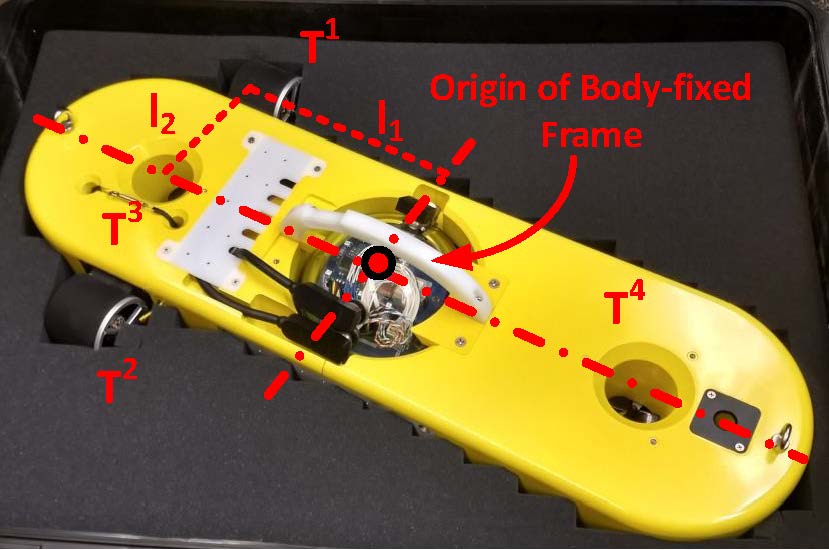} \vspace{-0.3cm}
\caption{Schematic of the DROP-Sphere}  \vspace{-0.5cm}
\label{fig:Sphere}
\end{figure}

DROP-Sphere, whose scheme is shown in Fig.\ref{fig:Sphere}, is propelled by four hub-less bi-directional thrusters powered by DC motors. Two horizontal thrusters are used for the surge and yaw controls, while two vertical thrusters are used for the heave and pitch controls. All the electrical devices (e.g., battery, camera, microprocessor) are mounted inside a transparent sphere located at the middle of the vehicle.  \vspace{-0.3cm}

\subsection{Mathematical Modeling}
\begin{figure}[!h]
\centering
\includegraphics[width=2.7in]{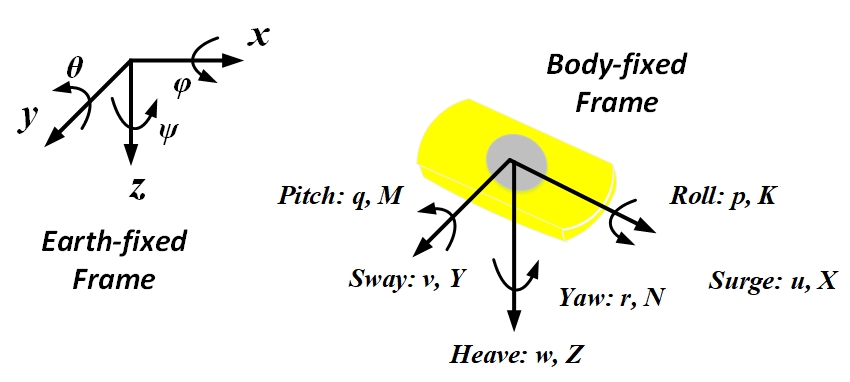} \vspace{-0.45cm}
\caption{Reference Frames and Notations}  \vspace{-0.5cm}
\label{fig:reference frames}
\end{figure}

To develop the equations of motion of an AUV, body-fixed and earth-fixed coordinate frames \cite{wadoo2017autonomous} are employed. Then the velocity, position, orientation, force and moment components (shown in Fig.~\ref{fig:reference frames}) are defined as follows:\vspace{-0.2cm}
\begin{equation} \vspace{-0.15cm}
\nu = \left[                 
  \begin{array}{cccccc}   
    u & v & w & p & q & r\\  
  \end{array}
\right] ^T,
\end{equation}
\begin{equation} \vspace{-0.15cm}
\eta = \left[                 
  \begin{array}{cccccc}   
    x & y & z & \phi & \theta & \psi \\  
  \end{array}
\right] ^T,
\end{equation}
\begin{equation} \vspace{-0.25cm}
\tau = \left[                 
  \begin{array}{cccccc}   
    X & Y & Z & K & M & N \\  
  \end{array}
\right] ^T,
\end{equation}
where, $\nu$, $\eta$, and $\tau$ are the vectors of velocities, positions and orientations, and external forces and moments respectively. \vspace{-0.45cm}

Based on the defined reference frames and notations, a general description of the nonlinear coupled AUV model is derived through the Newton-Euler equations of motion for a rigid body \cite{yuh1990modeling}. A compact form of the model is given as:
\vspace{-0.1cm}
\begin{equation}  \label{eq:Vehicle EOM} \vspace{-0.1cm}
\left\{  
             \begin{array}{lr}  
M_t\dot{\nu}+F_c(\nu)+F_h(\nu)\nu+F_g(\eta) = \tau_c,\\
\dot{\eta} = J(\eta)\nu,
             \end{array}  
\right.  
\end{equation}  
%
%where, $M_t$ is the vehicle total mass, $F_c(\nu)$ is the Coriolis and centripetal force, $F_h(\nu)$ is the hydrodynamic damping, $F_g(\eta)$ is the hydrostatic force, $\tau_c$ is the vector of control inputs, and $J(\eta)$ is the coordinate transformation between the body-fixed and earth-fixed frames.
where, $M_t$ is the total mass, $F_h(\nu)$ is the hydrodynamic damping matrix, $\tau_c$ is the control inputs.  The $F_c(\nu)$, $F_g(\eta)$ and $J(\eta)$ detailed in \cite{prestero2001verification} are the Coriolis and centripetal force, the hydrostatic force, and the coordinate transformation between the body-fixed and earth-fixed frames.

The total mass consists of the rigid body mass and the added mass. For the sake of simplicity and without consequential loss of accuracy, only diagonal terms in the added mass matrix ($M_a$) are considered: \vspace{-0.15cm}
%
% \begin{equation} \vspace{-0.1cm}
% M_t = M_{rb}+M_a,
% \end{equation}
\begin{equation} \vspace{-0.15cm}
M_a= -diag(X_{\dot{u}},Y_{\dot{v}},Z_{\dot{w}},K_{\dot{p}},M_{\dot{q}},N_{\dot{r}}),
\end{equation}
where, %$M_{rb}$ and $M_a$ are the rigid body mass and added mass. 
$X_{\dot{u}}$, $Y_{\dot{v}}$, $Z_{\dot{w}}$, $K_{\dot{p}}$, $M_{\dot{q}}$, $N_{\dot{r}}$ are the added mass coefficients. %The hydrodynamic damping matrix consists of hydrodynamic drag coefficients. 
Similarly, only diagonal and quadratic drag terms in hydrodynamic damping matrix are considered:\vspace{-0.45cm}

\begin{footnotesize} 
\begin{equation} \label{eq:Hydrodynamic Damping Matrix} \vspace{-0.15cm}
F_h(\nu) = -diag(X_{|u|u},Y_{|v|v}, Z_{|w|w}, K_{|p|p}, M_{|q|q}, N_{|r|r}) \cdot |\nu|,
\end{equation}
\end{footnotesize}%
where, $X_{|u|u}$, $Y_{|v|v}$, $Z_{|w|w}$, $K_{|p|p}$, $M_{|q|q}$, $N_{|r|r}$ are the quadratic drag coefficients. 
%The matrices of total mass, the Coriolis and centripetal force, and the hydrostatic force are given in Appendix~\ref{SecondAppendix}. 

% The coordinate transformation matrix $J(\eta)$ is in the block-diagonal form. $J(\eta)$ contains $J_1(\eta)$ and $J_2(\eta)$, which respectively relates translational and rotational velocities in the body-fixed frame to the positions and orientations in the earth-fixed frame: \vspace{-0.2cm}
% %
% \begin{equation}
% J(\eta) = \left[                 
%   \begin{array}{cc}   
%    J_1(\eta) &  0_{3\times3}\\  
%     0_{3\times3} & J_2(\eta)\\
%   \end{array}
% \right] .
% \end{equation}

% \begin{footnotesize}
% \begin{equation}
% J_1(\eta) = \left[                 
%   \begin{array}{ccc}   
%     c\psi c\theta & -s\psi c\phi+c\psi s\theta s\phi & s\psi s\phi+ c\psi s\theta c\phi\\  
%     s\psi c\theta & c\psi c\theta+s\psi s\theta s\phi & -c\psi s\phi+s\psi s\theta c\phi\\
%     -s\theta & c\theta s\phi & c\theta c\phi\\
%   \end{array}
% \right] ,
% \end{equation}
% \end{footnotesize}
% \begin{equation} \vspace{-0.1cm}
% J_2(\eta) = \left[                 
%   \begin{array}{ccc}   
%     1 & s\phi t\theta & c\phi t\theta\\  
%     0 & c\phi & -s\phi\\
%     0 & s\phi / c\theta & c\phi / c\theta\\
%   \end{array}
% \right] ,
% \end{equation} 
% %
% where, $c$, $s$, and $t$ are $cos(\cdot)$, $sin(\cdot)$, and $tan(\cdot)$, respectively. 

The control input vector $\tau_c$ consists of six components representing the control forces or moments in each DOF in the body-fixed frame. By combining the input transformation matrix with the thruster inputs, $\tau_c$ can be expressed as: \vspace{-0.4cm}

\begin{footnotesize}
\begin{equation}  \vspace{-0.2cm}
\setlength{\arraycolsep}{2pt}
\begin{split}
\tau_c &= \left[                 
  \begin{array}{cccccc}   
    \tau_X & \tau_Y & \tau_Z & \tau_K &  \tau_M & \tau_N\\ 
  \end{array}
\right] ^T \\
&=\left[                 
  \begin{array}{cccccc}   
    T^1+T^2 & 0 & T^3+T^4 & 0 &  l_1(T^3-T^4) & l_2(T^1-T^2)\\  
  \end{array}
\right] ^T,
\end{split}
\end{equation} 
\end{footnotesize}% 
where, as shown in Fig.~\ref{fig:Sphere}, $T^i$ is the input force of the $i^{th}$ thruster. $l_1$ is the distance between the vertical thrusters and the midship. $l_2$ is the distance from the horizontal thrusters to the center line.

%$l_1$, $l_2$ and $l_3$ are the distance between vertical thrusters and midship,  horizontal thrusters and center line, and horizontal thrusters and center of gravity.

The thruster inputs are converted from force to power in order to facilitate the energy consumption analysis. The power consumption of unit input force is calculated according to the momentum conservation theory as \cite{timmurphy.org}: \vspace{-0.15cm}
\begin{equation} \vspace{-0.2cm}
P(T^i) = \sqrt{\frac{1}{2\pi\rho}} \cdot \frac{(T^i)^{1.5}}{R} = C_p(T^i)^{1.5},
\end{equation}
in which, %$T^i$ is the input force of the $i^{th}$ thruster, 
$P(T^i)$ is the power consumption of the $i^{th}$ thruster, $\rho$ is the water density, $R$ is the radius of thruster, and $C_p$ is defined as the power conversion ratio. The numerical values for the vehicle parameters are provided in Appendix~\ref{FirstAppendix}.  

%%%%%%%%%%%%%%%%%%%%%%%%%%%%%%%%%%%%%%%%%%%%%%%%%%%%%%%%%%%%%%%%%%%%%%%%%%%%%%%%%%%%
%%%%%%%%%%%%%%%%%%%%%%%%%%%%%%%%%%%%%%%%%%%%%%%%%%%%%%%%%%%%%%%%%%%%%%%%%%%%%%%%%%%%
\section{Surge Motion Optimization Formulation} \label{section.3} 

%In this section, we propose a reformulation of the energy management problem based on the analysis of the global optimal solution to reduce the computational complexity.
%In this section, we propose an optimization based energy management strategy that is feasible for real-time application. \vspace{-0.1cm}
%First, the power management problem is formulated. Next, the optimal solution of the energy management problem is obtained via a global optimization method and it serves as the baseline for evaluating performance in the subsequent development. Finally, the energy management problem is reformulated as a surge motion optimization problem based on the analysis of global optimal solution to reduce the computational complexity. \vspace{-0.1cm}

\subsection{Energy Management Problem Formulation} 
In this paper, we consider energy-optimal maneuvering of the DROP-Sphere between two horizontal waypoints in an obstacle-free underwater area with no ocean currents. The vehicle is located at the initial position ($x_0$) heading towards the final position ($x_f$) with zero initial velocity in sway, heave, roll, pitch and yaw. While the controller is designed to drive the vehicle from $x_0$ to $x_f$, the energy consumed by the thruster input sequence $\{T^i_k\}$ is defined as: \vspace{-0.2cm}
\begin{equation} \label{eq:Eenergy Management Cost} \vspace{-0.15cm}
J(x_f-x_0,u_0,\{T^i_k\}) = \sum_{k=0}^{n-1}(\sum_{i=1}^4P(T^i_k)\cdot \frac{t_{travel}}{n}),
\end{equation}
with $x_f-x_0$ being the total distance, $k$ the index for step time, $n$ the number of samples, and $t_{travel}$ the travel time. 

The minimization of $J$ in (\ref{eq:Eenergy Management Cost}) is subject to (i) 6 DOF system dynamics described in (\ref{eq:Vehicle EOM}), and (ii) the state and input constraints given by the inequalities: \vspace{-0.15cm}
% %
% \begin{equation} \label{eq:Constraints for EM Problem} \vspace{-0.15cm}
% \begin{split}
% T_{min} \le  T^i_k \le  T_{max},\quad y_{min} \le y_k \le y_{max}, \\
% z_{min} \le z_k \le z_{max},\quad  \phi_{min} \le \phi_k \le \phi_{max} ,\\
% \theta_{min} \le \theta_k \le \theta_{max},\quad  \psi_{min} \le \psi_k \le \psi_{max},
% \end{split}
% \end{equation}
% %
%
\begin{equation} \label{eq:Constraints for EM Problem} \vspace{-0.15cm}
C_{min} \le  C_k \le  C_{max},
\end{equation}
where, $(\cdot)_{min}$ and $(\cdot)_{max}$ are the minimum and maximum of the corresponding variables, $C_k = [T^i_k, y_k, z_k, \phi_k, \theta_k, \psi_k]^T$.
\vspace{-0.1cm}
\subsection{Optimal Solution of the Energy Management Problem} 
%
%Treated as a trajectory optimization problem, the energy management problem in (\ref{eq:Eenergy Management Cost}) can be solved using various numerical methods. Direct collocation (DC), a direct method for numerically solving optimal control problems \cite{von1992direct}, is employed for deriving the global optimal solution. 
Treated as a trajectory optimization problem, the problem in (\ref{eq:Eenergy Management Cost}) can be solved with direct collocation (DC) \cite{von1992direct} or other numerical methods to obtain the global optimal solution. In this study, we consider the maneuvering from $x_0=0$ to $x_f=10$ with $u_0=0$. The state equation is discretized into $300$ segments and approximated with the trapezoid rule {to guarantee the optimality and robustness}.
%\textcolor{blue!50}{(I don't understand how the discretization level leads to global optimal solution? By using smaller grid size, you would reduce the numerical errors.)}
%%%%%%%%% this is the response to the first comment of reviewer 5
%The constraints enforced are $T\in[-7.86, 7.86]$, $y\in[-0.01, 0.01]$, $z\in[-0.005, 0.005]$, $\phi\in[-0.2, 0.2]$, $\theta\in[-0.01, 0.01]$, $\psi\in[-0.01, 0.01]$. 
The constraints are set as $C_{max} = -C_{min} = [7.84, 0.01, 0.005, 0.2, 0.01, 0.01]^T$. The units for the variables are $N$ for thrusts, $m$ for positions and $rad$ for orientations. 

\begin{figure}[!ht]
\centering
\includegraphics[width=3.5in]{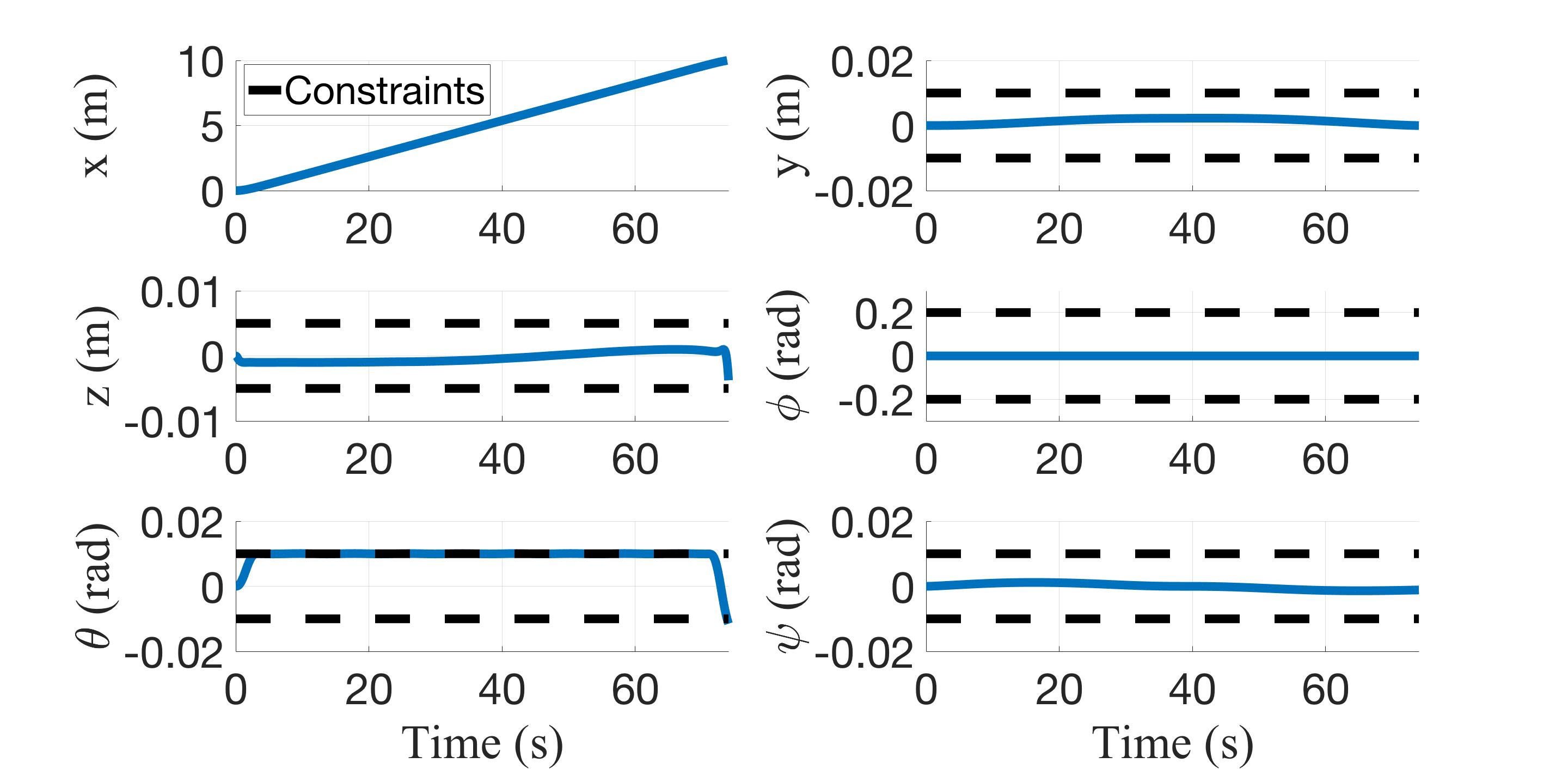} \vspace{-0.75cm}
\caption{Vehicle Trajectories from Direct Collocation Method} \vspace{-0.7cm}\label{fig:Trajectory of DC} 
\end{figure}

As illustrated by the resulted position and orientation traces shown in Fig.~\ref{fig:Trajectory of DC}, the optimal solution achieves point-to-point motion control while enforcing the constraints. However, DC method suffers from two issues. The open-loop nature of the solution makes it sensitive to modeling errors. In addition, their intensive computation renders the real-time implementation impractical. For example, solving (\ref{eq:Eenergy Management Cost}) with DC method for $74.04$~$s$ of simulation takes more than 3 hours on a 2.9 GHz Intel Core i5 processor with 8GB RAM. \vspace{-0.1cm}

\subsection{Optimization Problem Reformulation} 
To gain insights into the optimal operation of the DROP-Sphere, an in-depth analysis of the optimal solution %obtained via the DC method 
is conducted. %The total energy consumption is shown in Fig.~\ref{fig:Energy Allocation of DC}. %
{The total energy consumption ($69.08 \,J$) resulted from the DC method is distributed for surge ($22.08 \,J$), heave ($46.28 \,J$), pitch ($0.16\,J$) and yaw ($0.01\,J$) controls. Note that the energy used for controlling yaw ($0.01 \, \%$) and pitch ($0.23 \,\%$) are negligible compared to that of surge ($32.92 \, \%$) and heave ($66.83 \, \%$) controls.} 
%%%%%%%%% this is the response to the third minor comment of reviewer 7  
Given the constraints in (\ref{eq:Constraints for EM Problem}), then the yaw and pitch angles will be nearly zero during the whole trajectory. Thus, under this point-to-point motion control problem formulation, the energy saving potential from yaw and pitch controls is minimum. 
%\vspace{-0.3cm}
%
%\begin{figure}[!ht]
%\centering
%\includegraphics[width=3in]{Energy_allocation_DC.png} \vspace{-0.3cm}
%\caption{Energy Consumption Distribution of the Optimal Solution} \label{fig:Energy Allocation of DC} \vspace{-0.4cm}
%\end{figure}
%

\begin{figure}[!ht]
\centering
\includegraphics[width=3in]{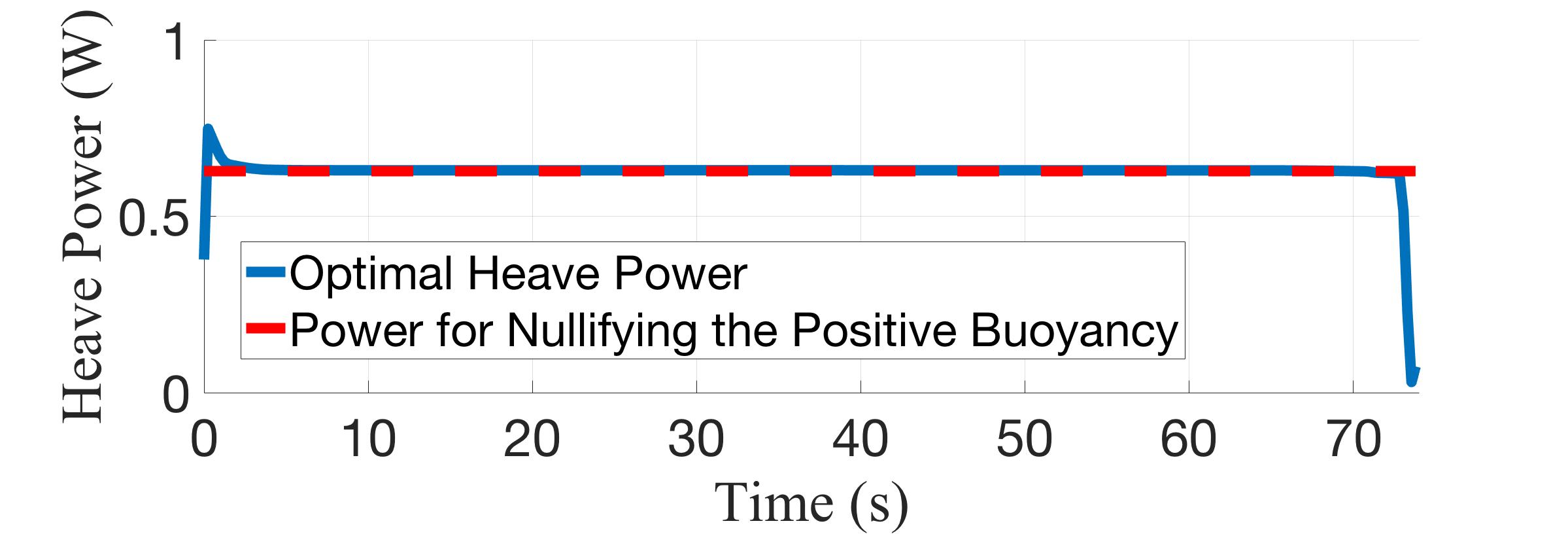}  \vspace{-0.35cm}
\caption{Heave Power Time History of the Optimal Solution} \vspace{-0.4cm}
\label{fig:Heave Power of DC} 
\end{figure}

Moreover, it can be observed from Fig.~\ref{fig:Heave Power of DC} that the heave power stays constant for most of the simulation time. By inspecting the heave dynamics in (\ref{eq:Vehicle EOM}), this constant value equals the power used for balancing the difference between the buoyancy and weight of the vehicle, i.e., the positive buoyancy. Given the pitch moment is approximately zero, the heave power can be calculated as: \vspace{-0.15cm}
\begin{equation} \label{eq:Heave power} \vspace{-0.2cm}
P^{PB} = \frac{\sqrt{2}}{2} C_p(B-W)^{1.5},
\end{equation}
where, $W$ and $B$ are the weight and buoyancy. Since the heave power is almost constant, the total heave energy becomes a linear function of the traveling time. This means that the heave energy is a reciprocal function of the surge speed, if the total distance is fixed. Thus, the energy saving could be substantial by increasing the surge velocity. On the other hand, higher surge speed leads to larger surge drag force ($X_{|u|u}|u|u$), which ultimately results in higher surge energy consumption. Thus, an optimization using the surge speed as the key variable can properly capture the sensitivity and trade-off of energy consumption to vehicle operation. \vspace{-0.2cm}

The analysis of the optimal solution leads to a decentralized control architecture where the yaw, pitch, and heave are independently controlled by three PID controllers and the energy management problem in (\ref{eq:Eenergy Management Cost}) is reformulated as a surge motion optimization problem. This results in a much simpler problem formulation compared to the original energy management problem. The surge motion optimization aims at finding the total horizontal thruster input sequence $\{T^{total}_k\}$ to minimize the following cost function: %\vspace{-0.5cm}
\begin{equation} \label{eq:Surge Velocity Optimization} \vspace{-0.25cm}
\begin{split}
J_{surge}(&x_f-x_0,u_0,\{T^{total}_k\}) = \\
&\sum_{k=0}^{n-1}(2 \cdot P(\frac{1}{2}T^{total}_k) +P^{PB})\cdot \frac{t_{travel}}{n},
\end{split}
\end{equation}
%
% %
% \begin{footnotesize}
% \begin{equation} \label{eq:Surge Velocity Optimization} \vspace{-0.25cm}
% J_{surge}(x_f-x_0,u_0,\{T^{total}_k\}) = \sum_{k=0}^{n-1}(2 \cdot P(\frac{1}{2}T^{total}_k)\cdot \frac{t_{travel}}{n}),
% \end{equation}
% \end{footnotesize}%
where, $T^{total}_k$ equals the combination of two horizontal thruster forces, which are same in magnitude and direction when the yaw moment is nearly zero. The optimization of $J_{surge}$ is subject to (i) input and state ($x_f$, $x_0$, $u_0$) constraints, and (ii) %2 DOF system dynamics (decoupled surge dynamic model) described as: 
{2 DOF dynamically decoupled surge dynamics (i.e., the accelerations from other DOF are considered as zero, while the velocities and orientations are assumed to be constant over the prediction horizon) described as:} \vspace{-0.15cm}
%%%%%%%%% this is the response to the third major comment of reviewer 7
%
\begin{equation} \label{eq:Surge Dynamics}
\begin{split}
(m-X_{\dot{u}})\dot{u} +m(wq-vr+&z_gpr)+(W-B)s\theta \\
&= -X_{|u|u}|u|u+T^{total},
\end{split}
\end{equation} \vspace{-0.3cm}
\begin{equation} \label{eq:Surge Kinematics} \vspace{-0.15cm}
\begin{split}
\dot{x} = (c\psi c\theta)u + (c\psi s\theta s\phi &-s\psi c\phi)v\\
&+ (s\psi s\phi+ c\psi s\theta c\phi)w, 
\end{split} 
\end{equation} 
where, $z_g$ is the center of gravity in the heave direction, $c$ and $s$ are $cos(\cdot)$ and $sin(\cdot)$ respectively.
%%%%%%%%%%%%%%%%%%%%%%%%%%%%%%%%%%%%%%%%%%%%%%%%%%%%%%%%%%%%%%%%%%%%%%%%%%%%%%%%%%%%
%%%%%%%%%%%%%%%%%%%%%%%%%%%%%%%%%%%%%%%%%%%%%%%%%%%%%%%%%%%%%%%%%%%%%%%%%%%%%%%%%%%%

\section{Static Surge Motion Optimization} \label{section.4}

One can approach the optimization problem defined in (\ref{eq:Surge Velocity Optimization}) by tracking an optimal surge velocity setpoint derived from solving a static optimization problem (i.e., assuming constant surge velocity). %Note that if the surge speed is constant, the total horizontal thrust ($T^{total}$) will be equal to the surge drag force ($X_{|u|u}|u|u$) in (\ref{eq:Surge Dynamics}). 
{Note that from the optimal solution, variations in the responses of the state variables other than surge velocity and $x$ position is minor. If we further assume that the surge speed is constant, then the total horizontal thrust ($T^{total}$) will be equal to the surge drag force ($X_{|u|u}|u|u$) in (\ref{eq:Surge Dynamics}).}
%%%%%%%%% this is the response to the third major comment of reviewer 7
According to the heave power expression in (\ref{eq:Heave power}), the total energy consumption can be described as: \vspace{-0.2cm}
\begin{equation} \label{eq:Terminal cost} \vspace{-0.2cm}
\begin{split}
J_{static}(x_f-x_0,u_0) = 
(x_f-x_0)&\cdot EPD(u_0),
\end{split}
\end{equation}
where, $EPD$ is \textit{energy cost per distance} defined as: \vspace{-0.1cm}
\begin{equation} \label{eq:Energy per distance} \vspace{-0.1cm}
EPD(u) = \frac{\sqrt{2}}{2} C_p X_{u|u|}^{1.5} u^2+ \frac{P^{PB}}{u}.
\end{equation}
The absolute operator on the surge velocity in $X_{|u|u}|u|u$ is dropped, as the negative surge velocity is not considered in this study. Based on (\ref{eq:Terminal cost}), the minimization of total energy will be equivalent to the minimization of $EPD$ in (\ref{eq:Energy per distance}) for given $x_f-x_0$. Since $EPD$ consists of a reciprocal function and a quadratic function of surge speed, there is a static optimal surge velocity ($u^*_{static}$) minimizing $EPD$ as well as the total energy consumption: \vspace{-0.2cm}
\begin{equation} \vspace{-0.15cm}
u^*_{static} = \mathop{\arg\min}_{u}\{EPD(u)\},
\end{equation} 
which represents the best trade-off between the energies for overcoming the positive buoyancy and the surge drag force.

To track the calculated $u^*_{static}$, a setpoint tracking MPC (T-MPC) is designed. The T-MPC calculates the total horizontal thruster input sequence $\{T^{total}_{k|t}\}$ over a prediction horizon $N$ to minimize the following cost function: \vspace{-0.15cm}
\begin{equation} \label{eq:Static MPC Cost} \vspace{-0.2cm}
J_{T}(u^*_{static},u_{t},\{T^{total}_{k|t}\}) = \sum_{k=0}^{N-1}(u^*_{static}-u_{k+1|t})^2,
\end{equation}
where, $(\cdot)_{k+1|t}$ is the $k$+$1$-step ahead prediction made at time $t$. The optimization in (\ref{eq:Static MPC Cost}) is subject to the decoupled surge dynamic model in (\ref{eq:Surge Dynamics}) and (\ref{eq:Surge Kinematics}), and the input constraints. 

\begin{figure}[!ht]
\centering
\includegraphics[width=3.5in]{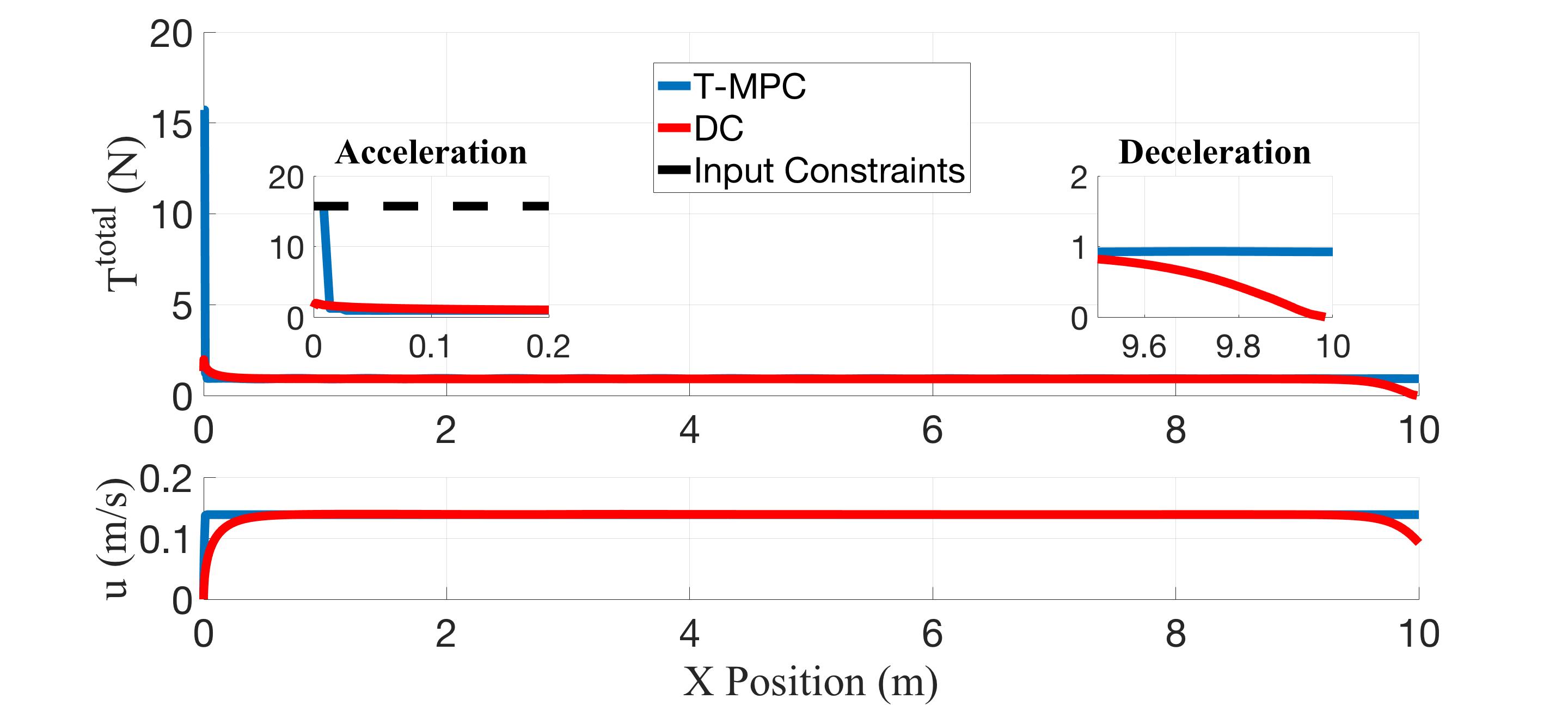}\vspace{-0.35cm}
\caption{Performance Comparison of T-MPC and DC} \vspace{-0.35cm}
\label{fig:Performance Comparison of T-MPC and DC}
\end{figure}

The T-MPC results in a sub-optimal performance in terms of energy consumption. For this case study, it results in $5.11\%$ additional energy cost compared to the optimal DC solution (i.e., $72.61J$ vs. $69.08J$). A comparison of surge velocity trajectory and total horizontal input from DC and T-MPC are shown in Fig.~\ref{fig:Performance Comparison of T-MPC and DC}. It can be seen from Fig.~\ref{fig:Performance Comparison of T-MPC and DC} that for most of the time, solution from T-MPC is close to that of DC with the same surge velocity, which verifies that tracking a statically optimized velocity setpoint can lead to ``almost'' optimal solution. However, during the acceleration and deceleration phases, the response of T-MPC differs from DC substantially. This is expected as T-MPC aims at reducing the tracking error regardless of the energy consumption. This motivates the dynamic surge motion optimization approach proposed in the next section to take the energy consumption into consideration during the dynamic stage. \vspace{-0.3cm}
%
%%%%%%%%%%%%%%%%%%%%%%%%%%%%%%%%%%%%%%%%%%%%%%%%%%%%%%%%%%%%%%%%%%%%%%%%%%%%%%%%%%%%
%%%%%%%%%%%%%%%%%%%%%%%%%%%%%%%%%%%%%%%%%%%%%%%%%%%%%%%%%%%%%%%%%%%%%%%%%%%%%%%%%%%%

\section{Dynamic Surge Motion Optimization } \label{section.5}
In order to incorporate energy-consciousness into the MPC, one can consider the following cost function: \vspace{-0.25cm}
\begin{equation} \label{eq:Stage Cost} \vspace{-0.15cm}
J_{L} (x_t,u_t,\{T^{total}_{k|t}\}) =\sum_{k=0}^{N-1}(2
\cdot P(\frac{1}{2} T^{total}_{k|t})+P^{PB}) \Delta t,
\end{equation}
where, $\Delta t$ is the sampling time. The policy to optimize (\ref{eq:Stage Cost}), however, will set the control input to zero for minimizing the energy consumption, unless the terminal position ($x_f$) is included in the MPC formulation as a state constraint and a long prediction horizon is used. With a long prediction horizon, the MPC becomes almost identical to those computationally demanding global optimization methods. 

To overcome this problem, we propose to add a terminal cost term ($J_{K}$) to $J_L$ in (\ref{eq:Stage Cost}) to reflect the ``cost-to-go'' and ensure that the MPC controller takes into account both the energy saving and the destination reaching. This terminal cost $J_{K}$ should approximate the ``cost-to-go'' and satisfy the following two requirements: (i) $J_{K}$ is an increasing function of the distance between $x_f$ and the $x$ position at the end of prediction horizon ($x_{N|t}$). (ii) $J_{K}$ provides an estimation of the energy consumption for reaching $x_f$ from $x_{N|t}$. 

Recalling the discussion in Section.~\ref{section.4}, we propose the following function as the terminal cost: \vspace{-0.2cm}
\begin{equation} \vspace{-0.2cm}
J_K = \frac{x_f-x_{N|t}}{u_{N|t}}\cdot (P^{PB}+\frac{\sqrt{2}}{2}C_p X_{u|u|}^{1.5}u_{N|t}^3),
\end{equation}
where, $u_{N|t}$ is surge velocity at the end of the prediction horizon. Then the overall cost function of the modified MPC called the energy-optimal MPC (EO-MPC) is described as: \vspace{-0.15cm}
\begin{equation} \label{eq:EO-MPC Cost} \vspace{-0.15cm}
J_{EO}(x_f-x_t,u_t,\{T^{total}_{k|t}\}) =J_L + J_K,
\end{equation}
where $J_{L}$ is defined in (\ref{eq:Stage Cost}). The optimization problem of (\ref{eq:EO-MPC Cost}) is subject to the same constraints as T-MPC. 

\begin{figure}[!h]
\centering
\includegraphics[width=3.3in]{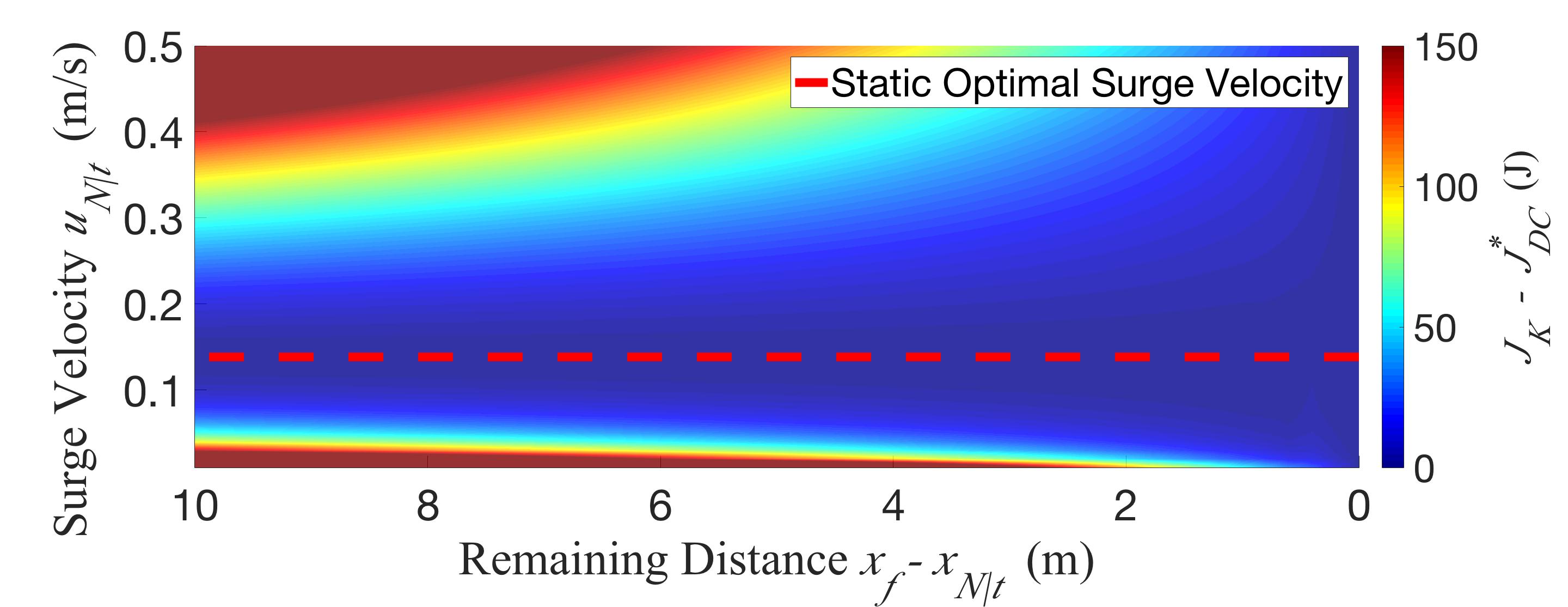} \vspace{-0.35cm}
\caption{$J_k - J^*_{DC}$ for Different Initial Conditions} \vspace{-0.4cm}
\label{fig:Kn_Jto_comp}
\end{figure}

According to the Bellman's Principle of Optimality~\cite{bellman2013dynamic},
%~\cite{lee2011model}, 
the EO-MPC would yield the optimal solution if $J_K$ were indeed the ``cost-to-go'', namely, $J_K(x_f-x_{N|t},u_{N|t})=J^*_{DC}(x_f-x_{N|t},u_{N|t},\{T^i_k\})$, where $J^*_{DC}$ is the global optimal solution from (\ref{eq:Eenergy Management Cost}). Extensive analysis has been carried out to evaluate the sensitivity of $J_K$ in estimating $J^*_{DC}$ with respect to different initial surge velocities at different distances to $x_f$. It can be seen from Fig.~\ref{fig:Kn_Jto_comp} that if the distance to $x_f$ is large, $J_K$ is close to $J^*_{DC}$ only when $u_{N|t}$ is near $u^*_{static}$. Thus, if the prediction horizon is tuned to ensure that (\ref{eq:Stage Cost}) is used for approximating the energy consumption in cost function during the region where the vehicle accelerates from $u_t$ to $u^*_{static}$, then a near-optimal energy consumption can be obtained with EO-MPC. \vspace{-0.3cm}

\begin{figure} [!h]
\centering
\includegraphics[width=3.2in]{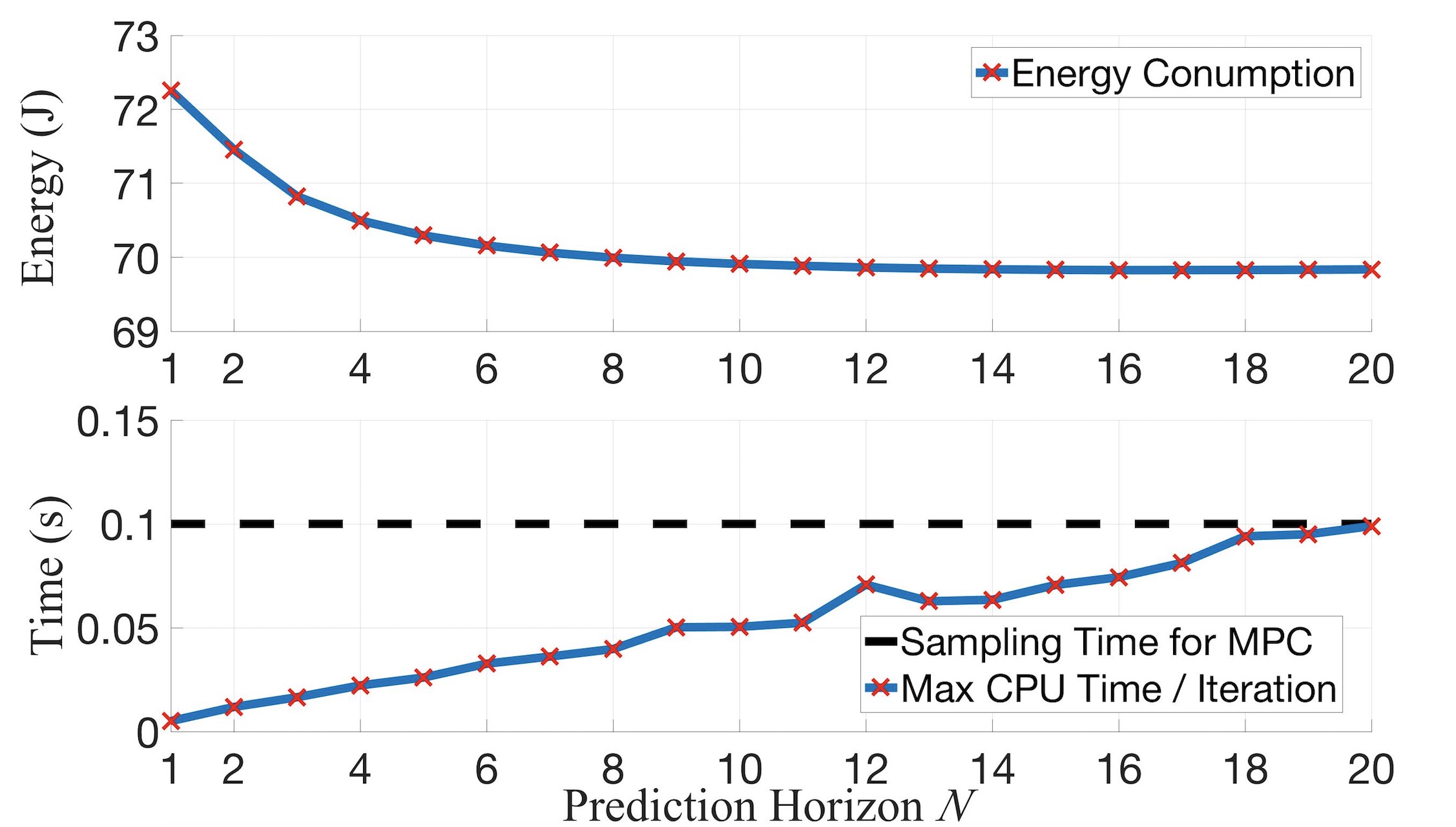}\vspace{-0.35cm}
\caption{Prediction Horizon Sensitivity Analysis} \vspace{-0.5cm}
\label{fig:Sensitivity of Prediction Horizon}
\end{figure}

Fig.~\ref{fig:Sensitivity of Prediction Horizon} shows the effect of the prediction horizon on energy consumption and the maximum CPU time for solving the optimization problem per iteration. It can be seen that as the length of the MPC prediction horizon increases, the energy consumption ultimately decreases and converges to $69.84J$, which is close to the energy consumption from the global optimal solution ($69.08J$). On the other hand, the CPU time increases as the prediction horizon becomes longer. Thus, in order to avoid high computational complexity and achieve near-optimal energy consumption, the prediction horizon can be selected as the minimum value that ensures the performance convergence~\cite{wojsznis2003practical}. Additionally, the maximum CPU time per iteration should also be smaller than the sampling time to guarantee the real-time execution of the MPC. For this case study, based on Fig.~\ref{fig:Sensitivity of Prediction Horizon}, the prediction horizon is chosen as $N=15$. It should be noted that because of the randomness associated with solving the optimization problem, the CPU time is calculated by taking average of the results from running the problem repeatedly for 10 times.  \vspace{-0.4cm}
%%%%%%%%%%%%%%%%%%%%%%%%%%%%%%%%%%%%%%%%%%%%%%%%%%%%%%%%%%%%%%%%%%%%%%%%%%%%%%%%%%%%
%%%%%%%%%%%%%%%%%%%%%%%%%%%%%%%%%%%%%%%%%%%%%%%%%%%%%%%%%%%%%%%%%%%%%%%%%%%%%%%%%%%%

\section{Real-time Energy-optimal MPC} \label{section.6}

\begin{figure}[!ht]
\centering
\includegraphics[width=0.9\columnwidth]{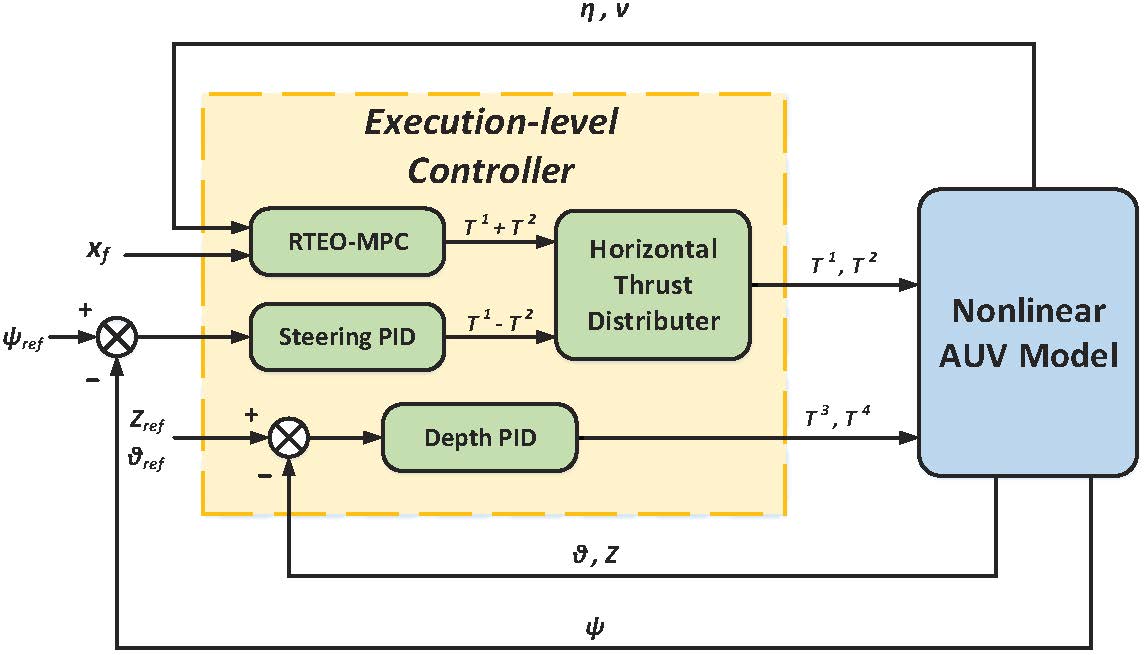} \vspace{-0.3cm}
\caption{Architecture of the Control System}
\label{fig:Architecture of the control system} \vspace{-0.4cm}
\end{figure}

With the proposed EO-MPC in Sec.~\ref{section.5}, the controller is able to achieve real-time near-optimal point-to-point motion control. However, as it was indicated from the performance comparison between T-MPC and DC, the benefit of conducting dynamic optimization in steady state is minimum. To further simplify the computational complexity of EO-MPC, a switching MPC algorithm is developed as the real-time energy-optimal MPC (RTEO-MPC). The RTEO-MPC takes different actions during the static and dynamic stages of the point-to-point motion control. As shown in {Algorithm~(1)}, 
%%%%%%%%% this is the response to the third comment of reviewer 3
at the beginning where the initial speed is different than the static optimal surge velocity ($u^*_{static}$) and at the end of the trajectory where the vehicle needs to decelerate in order to save energy, the RTEO-MPC has the same structure of the EO-MPC. During the cruise operation where the vehicle operates close to the $u^*_{static}$, the RTEO-MPC switches to a simple logic to keep the vehicle at a constant surge speed. The overall schematic of the real-time energy-optimal MPC (RTEO-MPC) along with the PIDs for depth and steering controls are illustrated in Fig.~\ref{fig:Architecture of the control system}. \vspace{-0.2cm}
\begin{algorithm}  
\caption{Switching Strategy for RTEO-MPC}\label{Algorithm:Switching Strategy}
\begin{algorithmic}[1]
\State Given the vehicle configuration, $x_f$, compute $u^*_{static}$;
\State Set $u^{switch}_{low}$, $u^{switch}_{high}$, $x^{switch}$;
\State \textbf{If} $x_t < x^{switch} $
\State \qquad \textbf{If} $u_0 < u^*_{static}$
\State \qquad \qquad \textbf{If} $u_t < u^{switch}_{low} \; or \; u_{t-1} < u_{t} $
\State \qquad \qquad \qquad  Compute $T^{total}_t$ using EO-MPC;
\State \qquad \qquad \textbf{Else}
\State \qquad \qquad \qquad  Set $T^{total}_t$ same as $T^{total}_{t-1}$;
\State \qquad \textbf{Else} 
\State \qquad \qquad \textbf{If} $u_t > u^{switch}_{high} \; or \; T^{total}_{t-2} < T^{total}_{t-1} $
\State \qquad \qquad \qquad  Compute $T_t^{total}$ using EO-MPC
\State \qquad \qquad \textbf{Else}
\State \qquad \qquad \qquad  Set $T^{total}_t$ same as $T^{total}_{t-1}$;
\State \textbf{Else}
\State \qquad Compute $T^{total}_t$ using EO-MPC;
\end{algorithmic} 
\algcomment{$Note$: $t$ is the current time. $u^{switch}_{low}$ and $u^{switch}_{high}$ are the low and high bounds around the $u^*_{static}$. $x^{switch}$ is the starting $X$ position after which the vehicle is near the destination.}
\end{algorithm}  \vspace{-0.6cm}
%%%%%%%%%%%%%%%%%%%%%%%%%%%%%%%%%%%%%%%%%%%%%%%%%%%%%%%%%%%%%%%%%%%%%%%%%%%%%%%%%%%%
%%%%%%%%%%%%%%%%%%%%%%%%%%%%%%%%%%%%%%%%%%%%%%%%%%%%%%%%%%%%%%%%%%%%%%%%%%%%%%%%%%%%

\section{Simulation Results And Analysis} \label{section.7}

This section presents the simulation results verifying the effectiveness of the proposed RTEO-MPC. In this case study, $x_0$, $u_0$, and $x_f$ are chosen as 0, 0, and 10. The thruster input constraint is considered as $-15.72 \, N \le T^{total} \le 15.72 \,N$. The sampling time is set as 0.1$s$. The MPC is implemented in MATLAB/Simulink on the full order model, as the virtual testbed, which provides full state feedback to the controllers. 

\begin{figure}[!h]
\centering
\includegraphics[width=3.3in]{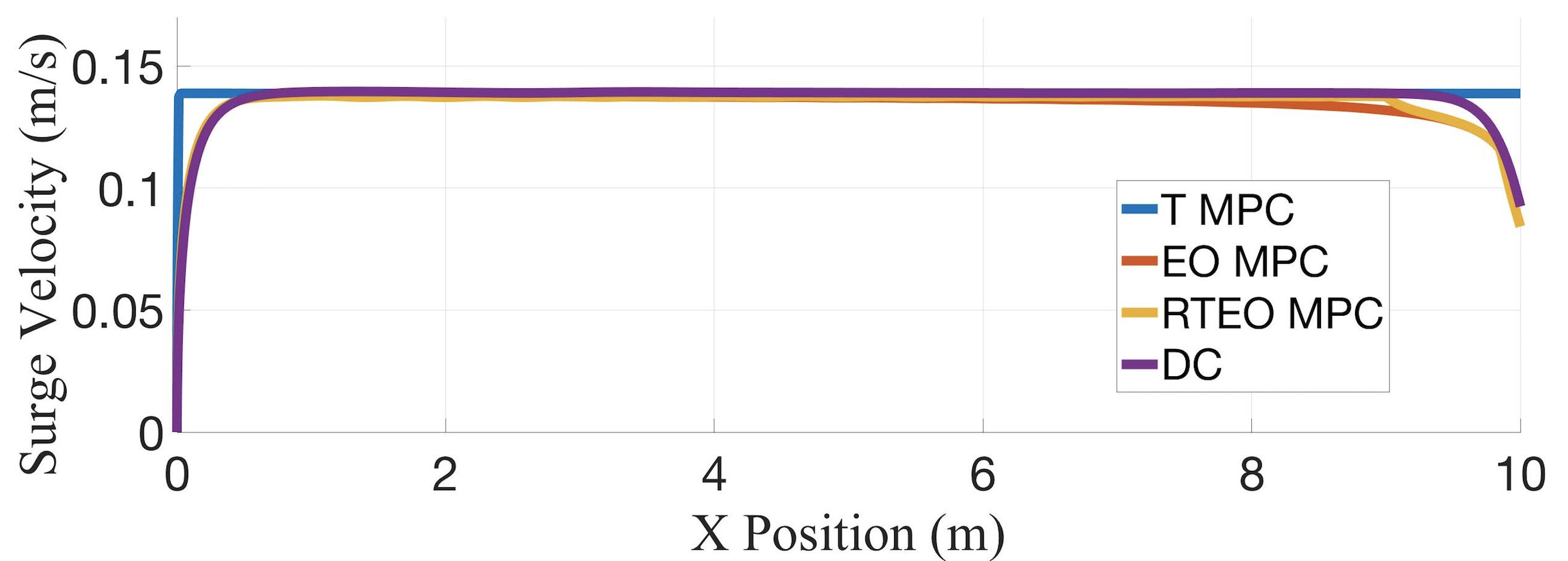} \vspace{-0.3cm}
\caption{Surge Velocity Trajectory Comparison} \vspace{-0.6cm}
\label{fig:Surge Velocity Profile}
\end{figure}

\begin{figure}[!h]
\centering
\includegraphics[width=3.5in]{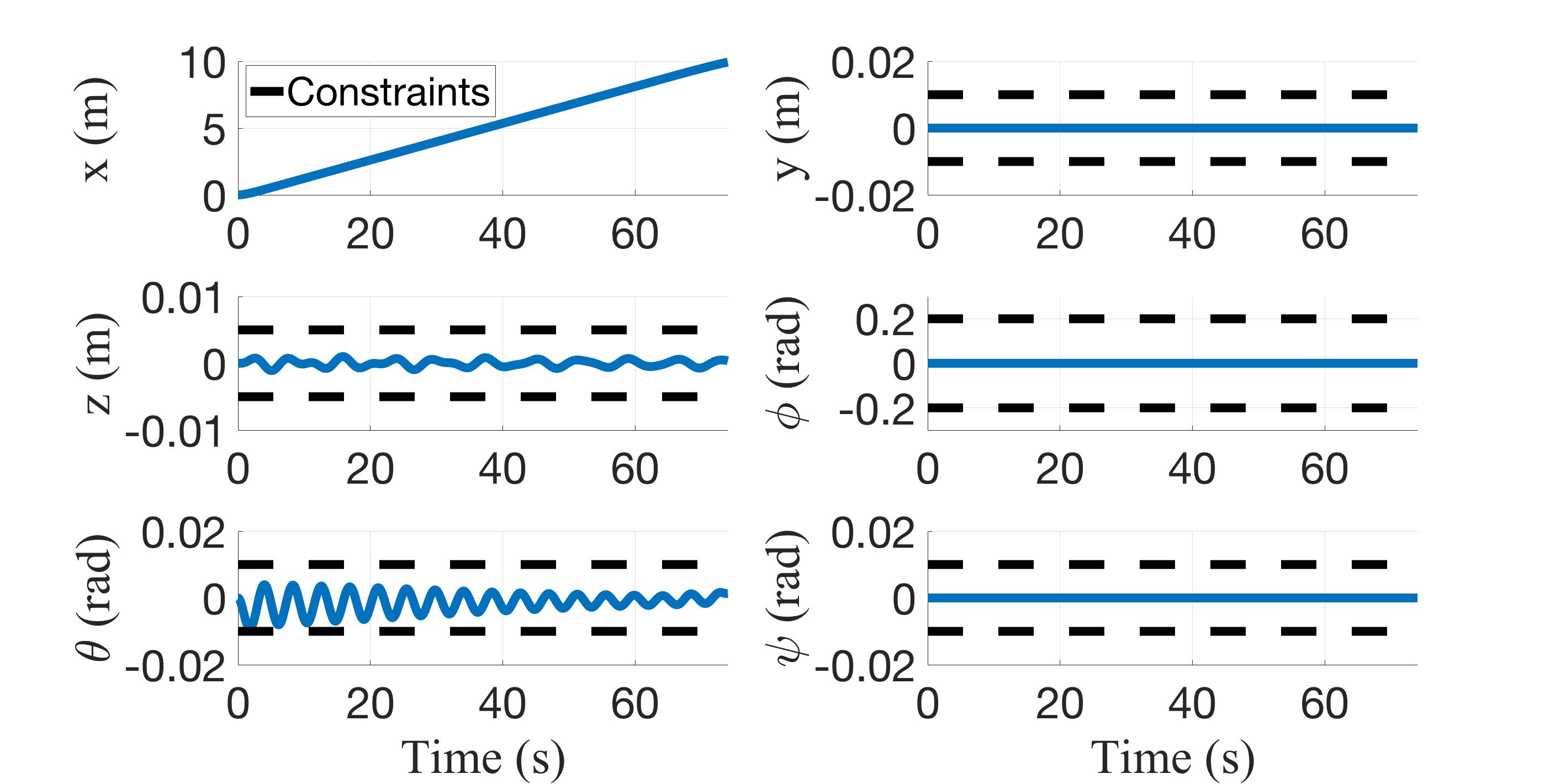} \vspace{-0.7cm} 
\caption{Vehicle Trajectories from the RTEO-MPC} \vspace{-0.4cm}
\label{fig:Vehicle Trajectory from RTEO-MPC}
\end{figure}

The energy consumption and the traveling time resulted from using the RTEO-MPC are listed in {Table~(I)}. 
%%%%%%%%%  this is the response to the third comment of reviewer 3
The performance of three other algorithms: (a) direct collocation (DC), (b) MPC for tracking velocity setpoint derived from static optimization (T-MPC), and (c) energy-optimal MPC (EO-MPC) are included for comparison. As indicated in {Table~(I),} 
%%%%%%%%%  this is the response to the third comment of reviewer 3
the performance of the proposed RTEO-MPC is close to the optimal baseline (DC) with $1.09\%$ more energy consumption. This difference is caused by assuming constant heave power in the cost function. Besides, compared to T-MPC, a $3.83\%$ reduction in energy cost is achieved with the RTEO-MPC. The RTEO-MPC improves the energy efficiency by leveraging the vehicle dynamics and optimizing the surge velocity trajectory during the dynamic stage, while, in the static stage, it guarantees the surge velocity to be close to the static optimal surge velocity ($u^*_{static}$) as shown in Fig.~\ref{fig:Surge Velocity Profile}. The time history of positions and orientations resulted from RTEO-MPC is illustrated in Fig.~\ref{fig:Vehicle Trajectory from RTEO-MPC}. It can be seen from Fig.~\ref{fig:Vehicle Trajectory from RTEO-MPC}, all the constraints enforced on the original energy management problem are satisfied by using the RTEO-MPC. \vspace{-0.7cm}
\newcommand{\tabincell}[2]{\begin{tabular}{@{}#1@{}}#2\end{tabular}} 
\begin{table}\footnotesize
\begin{spacing}{1.0}
\centering %\vspace{-0.2cm}
\caption{Performance Comparison} \vspace{-0.3cm}
\label{table:Performance Comparison}
\begin{tabular}{@{}cccc@{}}
\toprule
Control Strategy & Travel Time (s) &  \tabincell{c}{Energy \\Consumption (J)} & Loss (\%)  \\ \midrule
DC       &               74.04         &   69.08        &         --       \\ 
T-MPC       &          72.20              &    72.61       &           5.11           \\
EO-MPC       &              75.15          &    69.84       &        1.10             \\
RTEO-MPC     &              74.70          &   69.83        &          1.09           \\ \bottomrule
\end{tabular} 
\end{spacing} \vspace{-0.6cm}
\end{table} 

A further comparison of the computation time using two energy-optimal MPCs (EO-MPC and RTEO-MPC) and the DC method is presented in {Table~(II)}. 
%%%%%%%%%  this is the response to the third comment of reviewer 3
It can be seen that by using the switching strategy, the RTEO-MPC reduces the average computation time by $74.59\%$, without any deterioration in the energy efficiency compared to the EO-MPC. Moreover, compared to the DC method, the RTEO-MPC reduces the total computation time by $99.98\%$. %, which renders the RTEO-MPC real-time implementation feasible. %\vspace{-0.25cm}
\begin{table}[!h]\footnotesize
\begin{spacing}{1.0}
\centering
\caption{Computation Time Comparison}
\label{table:Computation Time Comparison}  \vspace{-0.25cm}
\begin{tabular}{@{}ccc@{}}
\toprule
Control Strategy & \tabincell{c}{Average CPU Time \\/ Sampling Time (s)}   & \tabincell{c}{Total CPU Time for \\ the Simulation of the Trip (s)} \\ \midrule
EO-MPC       &      0.0122           &      9.1824               \\ 
RTEO-MPC     &      0.0031          &        2.3503              \\ 
DC            &        --             &        11267.8                    \\\bottomrule
\end{tabular} \vspace{-0.6cm}
\end{spacing}
\end{table}

\begin{figure}[!h] 
\centering
\includegraphics[width=\columnwidth]{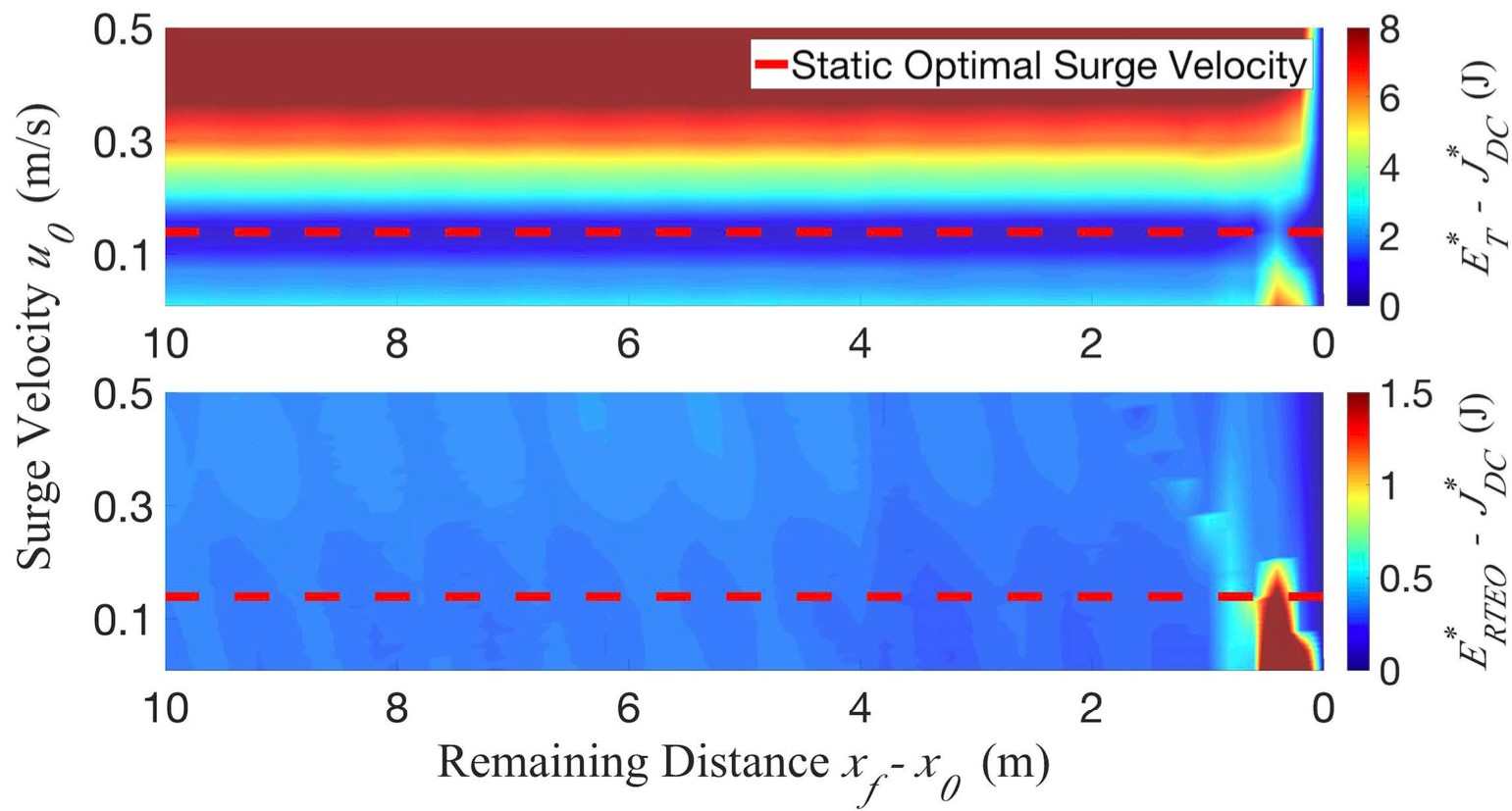} \vspace{-0.8cm}
\caption{$E^*_{T} - J^*_{DC}$ and $E^*_{RTEO} - J^*_{DC}$ for Different Initial Conditions} \vspace{-0.4cm}
\label{fig:Jmpc_Jto comp}
\end{figure}

To illustrate the robustness of the RTEO-MPC against different initial conditions, the energy consumption obtained from the RTEO-MPC ($E^*_{RTEO}$) and the T-MPC ($E^*_{T}$), with different initial positions and velocities, are compared to that of the DC method ($J^*_{DC}$) respectively, and the results are shown in Fig.~\ref{fig:Jmpc_Jto comp}. The ranges of initial position and velocity are set as $x_0\in [0,10]$ and $u_0\in [0,0.5]$. It can be seen from Fig.~\ref{fig:Jmpc_Jto comp} that the performance of the T-MPC deteriorates as the difference of initial velocity and the static optimal surge velocity increases, while the proposed RTEO-MPC is able to provide a near-optimal energy consumption under different initial conditions by combining dynamic and static optimizations via a switching logic. Higher energy consumption occurs for both the RTEO-MPC and the T-MPC when the vehicle starts very close to the target destination with low velocity, which is caused by neglecting the coupling effect of the heave and surge dynamics in the surge motion optimization problem formulation. However, this scenario is less likely to happen in an AUV deployment. \vspace{-0.3cm}
%%%%%%%%%%%%%%%%%%%%%%%%%%%%%%%%%%%%%%%%%%%%%%%%%%%%%%%%%%%%%%%%%%%%%%%%%%%%%%%%%%%%
%%%%%%%%%%%%%%%%%%%%%%%%%%%%%%%%%%%%%%%%%%%%%%%%%%%%%%%%%%%%%%%%%%%%%%%%%%%%%%%%%%%%
\section{CONCLUSION} \label{section.8}

In this paper, an innovative energy-optimal model predictive controller (MPC) was developed for real-time point-to-point motion control and energy management optimization of an AUV. To reduce the computational efforts and allow for real-time implementation of the proposed MPC, the overall AUV controller was designed and integrated based on a decentralized control architecture with the proposed MPC combined with several PIDs. The underlying physics of the vehicle motion and energy consumption is exploited by formulating the MPC cost function in terms of the total energy cost for transversing the remaining distance. Exploring the "static optimal" that achieves the best trade-off between energies for surge and heave controls is achieved, a switching logic was incorporated into the developed RTEO-MPC to switch the control law between static and dynamic stages of the surge motion for further reduction in the computation time. Simulation results verified that the proposed method allows for the real-time trajectory optimization with near-optimal energy consumption under different initial conditions. For a case study presented, the RTEO-MPC showed a $3.83\%$ improvement in energy consumption compared to the MPC for tracking static optimal surge velocity setpoint (T-MPC), and $99.98\%$ reduction in total computation time compared to the global optimization method (DC). 

%\section*{Future Works}

{Future works will be focusing on generalizing this framework to (i) complicated maneuvering operations (e.g., steering and diving), and (ii) realistic environmental conditions (e.g., ocean currents and obstacles). Meanwhile, theoretical results about the recursive feasibility and optimality resulted from the proposed terminal cost will be investigated for the point-to-point navigation scenario. }%\textcolor{blue!50}{this is not a strong argument, you cannot theoretically grantee something with some demonstration.}}
%%%%%%%%%  this is the response to the first/second/fourth major comment of reviewer 7

%The proposed RTEO-MPC was evaluated for a point-to-point AUV motion control ignoring the ocean currents and the vehicle turning at waypoints. Future works will be focusing on incorporating the real-time current information into the energy-consciousness AUV motion control and considering more complicated and general maneuvering operations.  

%\addtolength{\textheight}{-15cm}   % This command serves to balance the column lengths

%%%%%%%%%%%%%%%%%%%%%%%%%%%%%%%%%%%%%%%%%%%%%%%%%%%%%%%%%%%%%%%%%%%%%%%%%%%%%%%%
\section*{ACKNOWLEDGMENT}
The authors thank Dr. Corina Barbalata, and Mr. Eduardo Iscar R\"{u}land from DROP lab at the University of Michigan for providing details of the AUV model and their technical comments during this study.

%The authors thank Prof. Matthew Johnson-Roberson, Dr. Corina Barbalata, and Mr. Eduardo Iscar R\"{u}land from DROP lab at the University of Michigan for providing details of the AUV model and their technical comments during this study. 

%%%%%%%%%%%%%%%%%%%%%%%%%%%%%%%%%%%%%%%%%%%%%%%%%%%%%%%%%%%%%%%%%%%%%%%%%%%%%%%%

\bibliographystyle{IEEEtran}
\bibliography{Reference.bib}

%%%%%%%%%%%%%%%%%%%%%%%%%%%%%%%%%%%%%%%%%%%%%%%%%%%%%%%%%%%%%%%%%%%%%%%%%%%%%%%%
\begin{appendices}
% \section{Matrices in Vehicle Equations of Motion}
% \label{SecondAppendix} \vspace{-0.5cm}
% \begin{tiny}
% \begin{equation}
% \setlength{\arraycolsep}{1.5pt}
% M_t = \left[                 
%   \begin{array}{cccccc}   
%     m-X_{\dot{u}} & 0 & 0 & 0 & mZ_g & 0\\  
%     0 & m-Y_{\dot{v}} & 0 & -mZ_g & 0 & 0\\
%     0 & 0 & m-Z_{\dot{w}} & 0 & 0 & 0\\
%     0 & -mZ_g & 0 & I_{xx}-K_{\dot{p}} & 0 & 0\\
%     mZ_g & 0 & 0 & 0 & I_{yy}-M_{\dot{q}} & 0\\
%     0 & 0 & 0 & 0 & 0 & I_{zz}-N_{\dot{r}}\\
%   \end{array}
% \right] ,
% \end{equation}
% \end{tiny}
% \begin{equation}
% F_{c}(\nu) = \left[                 
%   \begin{array}{c}   
%     m(-vr+wq+z_gpr)\\  
%     m(-wp+ur+z_gqr)\\
%     m[-uq+vp-z_g(p^2+q^2)]\\
%     (I_{zz}-I_{yy})qr-mz_g(-wp+ur)\\
%     (I_{xx}-I_{zz})rp+mz_g(-vr+wq)\\
%     (I_{yy}-I_{xx})pq\\
%   \end{array}
% \right] ,
% \end{equation}
% \begin{equation}
% F_{g}(\eta) = \left[                 
%   \begin{array}{c}   
%     (W-B)sin\theta\\  
%     -(W-B)cos\theta sin\phi\\
%     -(W-B)cos\theta cos\phi\\
%     (z_gW - z_bB)cos\theta sin\phi\\
%     (z_gW - z_bB)sin\theta\\
%     0 \\
%   \end{array}
% \right] ,
% \end{equation}
% where $I_{xx}$, $I_{yy}$, $I_{zz}$ are vehicle's moment of inertia. 

\section{Sphere Vehicle Parameters}
\label{FirstAppendix}
%\begin{table}
\begin{scriptsize}
\centering
\label{my-label}
\begin{tabular}{lll}
$W=200.116 \, N$ & $B = 201.586\, N$ & $m = 20.42\, kg$\\
$I_{xx} = 0.1205 \,kg\, m^2$  &  $I_{yy} = 0.9431 \,kg\, m^2$  &   $I_{zz} = 1.0061\, kg\, m^2$\\
$z_g = 0.0018\, m$  &  $l_1 = 0.1694 \, m$            &   $l_2 = 0.2794 \, m$ \\
$R = 0.025 \, m$  &  $X_{\dot{u}} = -2.042 \, kg$           &     $Y_{\dot{v}} = -32.2013\, kg$   \\
$Z_{\dot{w}} = -32.2013\, kg$   &  $K_{\dot{p}} = -0.0805\, kg$           &     $M_{\dot{q}} = -2.6834\, kg$  \\
$N_{\dot{r}} = -2.6834\, kg$   &  $X_{u|u|} = 48.17 \, kg/m$           &     $Y_{v|v|} = 4.11\, kg/m$ \\
$Z_{w|w|} = 4.11\, kg/m$   &  $K_{p|p|} = 48.17\, kg/m$           &     $M_{q|q|} = 4.11\, kg/m$ \\
$N_{r|r|} = 4.11\, kg/m$   &  $\rho = 1.025 \,kg/m^3$           &    
\end{tabular}
%\end{table}
\end{scriptsize}
\end{appendices}

\end{document}